\documentclass[superscriptaddress,twocolumn,showpacs,prb]{revtex4-1}
\usepackage[utf8]{inputenc} 
\usepackage{amsmath}
\usepackage{braket}
\usepackage{graphicx}
\usepackage{amsfonts}
\usepackage{pgfplots}
\usepackage{csquotes}
\usepackage{hhline}
\usepackage{amssymb}
\usepackage{listings}
\usepackage{color}

\definecolor{codegreen}{rgb}{0,0.6,0}
\definecolor{codegray}{rgb}{0.5,0.5,0.5}
\definecolor{codepurple}{rgb}{0.58,0,0.82}
\definecolor{backcolour}{rgb}{0.95,0.95,0.92}
 
\lstdefinestyle{mystyle}{
    backgroundcolor=\color{backcolour},   
    commentstyle=\color{codegreen},
    keywordstyle=\color{magenta},
    numberstyle=\tiny\color{codegray},
    stringstyle=\color{codepurple},
    basicstyle=\footnotesize,
    breakatwhitespace=false,         
    breaklines=true,                 
    captionpos=b,                    
    keepspaces=true,                 
    numbers=left,                    
    numbersep=5pt,                  
    showspaces=false,                
    showstringspaces=false,
    showtabs=false,                  
    tabsize=2
}
 
\usepackage{subfigure}

\usepackage{dcolumn}
\usepackage{tabularx}
\usepackage[colorlinks=true,linkcolor=blue,citecolor=blue,urlcolor=blue]{hyperref}
\usepackage{longtable}
\usepackage{braket}
\usepackage{float}
\newcolumntype{C}{>{\centering\arraybackslash}X}
\begin{document}
\title{Masking of Quantum Information into Restricted Set of states}

\author{Tamal Ghosh}
\email{tamalghosh695@gmail.com}
\affiliation{Physics Department, \\West Bengal State University, Barasat 700126, West Bengal, India}
\author{Soumya Sarkar}
\email{sarkarsoumya65@gmail.com}
\affiliation{Department of Physics, \\ National Institute of Technology Karnataka, Surathkal 575025, Karnataka, India}

\author{Bikash K. Behera$^@$}
\thanks{$^@$Corresponding Author}
\email{bikash@bikashsquantum.com}
\affiliation{Bikash's Quantum (OPC) Pvt. Ltd., Balindi, Mohanpur 741246, West Bengal, India}
\affiliation{Department of Physical Sciences,\\ Indian Institute of Science Education and Research Kolkata, Mohanpur 741246, West Bengal, India}
\author{Prasanta K. Panigrahi}
\email{pprasanta@iiserkol.ac.in}
\affiliation{Department of Physical Sciences,\\ Indian Institute of Science Education and Research Kolkata, Mohanpur 741246, West Bengal, India}

\begin{abstract}
Masking of data is a method to protect information by shielding it from a third party, however keeping it usable for further usages like application development, building program extensions to name a few. Whereas it is possible for classical information encoded in composite quantum states to be completely masked from reduced sub-systems, it has to be checked if quantum information can also be masked when the future possibilities of a quantum computer are increasing day by day. Newly proposed no-masking theorem [Phys. Rev. Lett. \textbf{120}, 230501 (2018)], one of the no-go theorems, demands that except for some restricted sets of non-orthogonal states, it's impossible to mask arbitrary quantum states. Here, we explore the possibility of masking in the IBM quantum experience platform by designing the quantum circuits, and running them on the 5-qubit quantum computer. We choose two particular states considering both the orthogonal bipartite and a tripartite system and illustrate their masking through both the theoretical calculation as well as verification in the quantum computer. By quantum state tomography, it is concluded that the experimental results are collected with high fidelity and hence the possibility of masking is realized.
\end{abstract}

\begin{keywords}{No-Masking Theorem, IBM Quantum Experience}\end{keywords}
\maketitle
\section{Introduction \label{qnm_Sec1}}
Before talking about quantum information and it's security, let us first know briefly what classical information and it's security methods are, as classical world is more intuitive to us. 

Classical information was first carefully defined by Shannon in his paper in 1948 \cite{qnm_ShannonIEEE1948}. `Bit' is the basic unit of classical data. Though there is a fine and important difference between information and data (information is a special type of data that is not known already). In today's digital era, we use voltage to create states `0' and `1' as bits. Now there's a lot of data security techniques to make the data communication secured. Data encryption that transfers the actual data into an ineffable one that is useless to a hacker. By data encapsulation, one can only perform a restricted set of operations. Data anonymization makes an user anonymous while using the internet. Classical data masking replaces the actual data with a fictional one that somehow represents the production data but the third party never identifies it.

Let us now see how classical information can also be  encoded in a composite quantum system. Suppose, we encode a single bit classical information in two orthogonal entangled states. The mapping is as follows:

\begin{eqnarray}
\Ket{0}\xrightarrow{}\frac{1}{\sqrt{2}}\Big(\Ket{00}+\Ket{11}\Big) \nonumber\\
\Ket{1}\xrightarrow{}\frac{1}{\sqrt{2}}\Big(\Ket{00}-\Ket{11}\Big)
\label{qnm_Eq1}
\end{eqnarray}

Here, we can see that the subsystems do not have any information about the actual input classical bit. Hence, the information is masked. However, in quantum realm, the classical bits are analogous to `qubits'. Instead of just two definite states `0' and `1' in classical information scheme, a quantum computer uses the superposition of the both, making it more efficient in problem solving. On the other hand, the linearity and the unitarity make the quantum communication more secured. These result the set of no-go theorems \cite{qnm_OldofrediJGPS2018}. A pure quantum state can not be perfectly copied, known as the no-cloning theorem \cite{qnm_WottersNat1982}. Suppose, we have sub-systems $A$ and $B$ in Hilbert spaces $H_{A}$ and $H_{B}$. Now, we want to copy the initial state $\ket{\Phi}_{A}$ in the quantum sub-system $B$. To do that, we need a blank state $\Ket{k}$ which is independent of $\ket{\Phi}_{A}$. The theorem says, we do not have any operator $U$ on the Hilbert space $H_{A} \otimes H_{B}$ such that: $U\Ket{\Phi k}_{AB}\xrightarrow{}\Ket{\Phi\Phi}_{AB}$. No-Broadcasting theorem \cite{qnm_BarnumPRL2007,qnm_KalevPRL2008,qnm_MarvianPRL2019} tells that it is impossible to have a map $\epsilon$ that broadcasts the state $\rho$ from $H$ to $H_{A} \otimes H_{B}$ if $Tr_{A}(\epsilon(\rho))=Tr_{B}(\epsilon(\rho))$ - consequence of no-cloning theorem. No-hiding \cite{qnm_BraunsteinPRL2007} and no-deleting theorems \cite{qnm_PatiNat2000} support the quantum information conservation in our universe. The no-hiding theorem has been experimentally realized by using nuclear magnetic resonance \cite{qnm_SamalPRL2011} and then later got tested on the IBM quantum computer \cite{qnm_KalraQIP2019}. Unlike classical information, quantum information cannot be completely hidden in correlations between a pair of subsystems.

However, in those theorems, there are always some restricted conditions for which the theorems no longer hold. If we consider no-cloning theorem, in some cases imperfect clones can be produced if a larger auxiliary system is coupled with the original state and a perfect unitary operation is done on the combined system, then some components of the system evolve to approximate copies of the original state. No-broadcasting theorem can not be generalized to more than a single input copy. Even $\textbf{\textit{Superbroadcasting}}$ \cite{qnm_DArianoPRL2005} tells that it's even possible for four or more inputs to extract the input states while broadcasting. 

Recently, Modi \emph{et al.} proposed the scheme of masking of quantum information \cite{qnm_ModiPRL2018}, where they defined the masking conditions. They concluded that it is not possible to mask arbitrary quantum state, however, some restricted states of non-orthogonal quantum states can be masked. In the present work, we experimentally show that even though some arbitrary quantum information can not be masked, but like the above, no-masking theorem also does not hold always for some particular quantum states. Here, we choose particular two-qubit states with orthogonal basis sub-states and work out the no-masking theorem conditions. It is found that the above state satisfies the conditions of masking, and can be masked. The two-qubit quantum states are prepared on the IBM quantum experience platform, and the quantum circuits are designed, and run on the real quantum chip ``ibmqx4". For more experiments performed on this platform, the following works can be referred \cite{qnm_Rui2017,qnm_KapilarXiv2018,qnm_DasharXiv2018,qnm_SwainQIP2019,qnm_Unai2018,qnm_Debjit2018,qnm_He2017,qnm_Saipriya2018,qnm_Huang2017,qnm_Garc2018,qnm_Behera2019,qnm_Harper2019,qnm_Bikash2019}. The experimental results are collected, and compared with the theoretically predicted ones. From the quantum state tomography, it is observed that with more than 98\% fidelity, the expected results are obtained for the case of orthogonal basis states. To avoid confusion it should be clear to the readers that here we have used two types of fidelities - one, that tells how perfectly the states are prepared in IBM quantum computer by comparing the practical values to theoretical values and the second type tells the measurement of masking between the practical results of a given state. 

The organization of the paper is as follows. In Section \ref{qnm_Sec2}, we define the operation of quantum masking, then verify that masking of quantum information possible into some orthogonal state in Section \ref{qnm_Sec3} with a restricted condition. For extra support, in Section \ref{qnm_Sec4}, we create a quantum state with arbitrary coefficients with higher than 99\% fidelity of gate operations and it is seen that masking is not quite satisfactory compared to those restricted states and in \ref{qnm_Sec5} we verify an interesting fact which fails for bipartite system but not in the case of tripartite state. In Section\ref{qnm_Sec6} we study the statistical errors in our experimental method. Finally, we conclude in Section \ref{qnm_Sec7} discussing the experimental results.

\section{Definition\label{qnm_Sec2}}
According to the Ref. \cite{qnm_ModiPRL2018}, conditions for masking are defined as follows. Let us assume quantum information is in the states $\Ket{a_{k}}_{A}$ belonging to $H_{A}$. If there is an operator $M$ that maps the states into $\Ket{\Psi_{AB}}$ belonging to $H_{A} \otimes H_{B}$ such that it satisfies the following two conditions,

\begin{eqnarray}
\rho_{A}=Tr_{B}\big(|\Psi_{k}\rangle_{AB}\langle\Psi_{k}|\big), \nonumber\\
\rho_{B}=Tr_{A}\big(\Ket{\Psi_{k}}_{AB}\langle\Psi_{k}|\big)
\end{eqnarray}

are identical and one can say nothing about the value of $k$ by observing this. As this is a physical process, it can also be written as $M: \Ket{a_{k}}_{A} \otimes \Ket{b}_{B}\xrightarrow{}\Ket{\Psi_{k}}_{AB}$, where $M$ is called the masker and $U$ acts on both the system $A$ and $B$.

\section{Masking of quantum information into Orthogonal States\label{qnm_Sec3}}
Here, we consider a quantum state, and show that it can be decomposed in such way that this state can be masked. Hence, we illustrate that information can also be masked into orthogonal states, if they can be decomposed in a proper way.

We pick up only one of the very few states, which follows the masking condition theoretically (we show below how they are following the conditions of masking) among all the combinations of orthogonal states. Hence, we opt out other states which don't follow those conditions.
\subsection{Circuit Explanation}

\begin{figure}[H]
\centering
\includegraphics[scale=0.7]{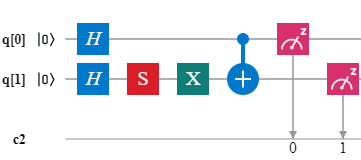}
\caption{Quantum circuit for generating $\Ket{\Psi}=\alpha_1\Ket{\Psi_0}+\alpha_2\Ket{\Psi_1}$, where, $\Ket{\Psi_{0}}=\Big(\frac{\Ket{00}+\Ket{11}}{\sqrt2} \Big)$, and $\Ket{\Psi_{1}}=\Big(\frac{\ket{01}+\Ket{10}}{\sqrt2} \Big)$.}
\label{qnm_Fig2}
\end{figure}

In this case, to construct the state which is linear combination of two mutual orthogonal sub-states $\ket{\Psi_0}$ and $\ket{\Psi_1}$, we use two Hadamard gates, one $S$-gate or phase gate, one $X$-gate, and one CNOT gate. The Pauli-$X$ gate acts on a single qubit, it is the quantum equivalent of the NOT gate for classical computers (with respect to the standard basis $\ket{0},\Ket{1}$, which distinguishes the Z-direction. It equates to a rotation around the X-axis of the Bloch sphere by $\pi$ radians. It maps $\Ket{0}$ to $\Ket{1}$ and $\Ket{1}$ to $\Ket{0}$. Due to this nature, it is sometimes called bit-flip gate. It is represented by the Pauli-X matrix. The phase gate (or $S$ gate) is a single-qubit operation, the $S$ gate is also known as the phase gate, because it represents a 90 degree rotation around the Z-axis. In the previous section, we have already discussed about the Hadamard and CNOT gate. 

\subsection{Theory}

In this section, we take an orthogonal quantum state and show theoretically that the reduced states are identical and the experiment supports the calculation as well. We now assume that $\Ket{b}$ can be masked, i.e.,

\begin{equation}
\Ket{b}=\alpha_1\Ket{0}+\alpha_2\Ket{1}\xrightarrow{}\Ket{\Psi}=\alpha_1\Ket{\Psi_0}+\alpha_2\Ket{\Psi_1}
\end{equation}
where $|\alpha_1|^2+|\alpha_2|^2=1$. Now, we take the partial traces with respect to either $A$ or $B$ to get

\begin{eqnarray}
Tr_{x}\Ket{\Psi}\langle\Psi|&=& \alpha_1^2Tr_{x}(\Ket{\Psi_0}\langle\Psi_0|)+\alpha_2^2Tr_x(\Ket{\Psi_1}\langle\Psi_1|)\nonumber\\&+&\alpha_1\alpha_2^*Tr_{x}(\Ket{\Psi_0}\langle\Psi_1|)+\alpha_1^*\alpha_2Tr_{x}(\Ket{\Psi_1}\langle\Psi_0|)\nonumber\\
\label{qnm_Eq4}
\end{eqnarray}

Now the masking condition is $\rho_y=Tr_{x}(\Ket{\Psi_0}\langle\Psi_0|)=Tr_{x}(\Ket{\Psi_1}\langle\Psi_1|)$. To fulfill the masking condition, the off-diagonal terms in the Eq. \eqref{qnm_Eq4} must be vanished. So,

\begin{eqnarray}
\alpha_1\alpha_2^*Tr_{x}(\Ket{\Psi_0}\langle\Psi_1|)+\alpha_1^*\alpha_2Tr_{x}(\Ket{\Psi_1}\langle\Psi_0|)=0
\label{qnm_Eq5}
\end{eqnarray}

Now, from the calculations done in Eq. \eqref{qnm_Eqn12} and Eq. \eqref{qnm_Eqn13}, it can be seen that $Tr_{X}\Ket{\Psi_{1}}\big\langle\Psi_{0}\big|$ and $Tr_{X}\Ket{\Psi_{0}}\big\langle\Psi_{1}\big|$, these are non-zero and equal terms, hence Eq. \eqref{qnm_Eq5} can be written as

\begin{eqnarray}
Tr_{x}\Ket{\Psi_0}\langle\Psi_1|(\alpha_1\alpha_2^*+\alpha_1^*\alpha_2)=0 \nonumber\\
\alpha_1\alpha_2^*+\alpha_1^*\alpha_2=0\label{qnm_Eqn5}
\end{eqnarray}

when

\begin{eqnarray}
Tr_{X}\Ket{\Psi_{1}}\big\langle\Psi_{0}\big|=Tr_{X}\Ket{\Psi_{0}}\big\langle\Psi_{1}\big|
\label{qnm_Eqn6}
\end{eqnarray}

The above condition means that we have to choose $\alpha_1$ and $\alpha_2$ in such way so that one is imaginary part of another. \textit{\textbf{As $\alpha_1$ and $\alpha_2$ are the information we want to mask, so with this condition we impose a restriction on the information we want to mask}}. It is important to note that along with the states (which we are masking the information into) which must follow the masking condition, the quantum information, itself, needs follow the above condition to get masked.

Here we take a state $\Ket{\Psi}$ which is a linear combination of two orthogonal states,

\begin{eqnarray}
\Ket{\Psi_0}=\Big(\frac{\Ket{00}+\Ket{11}}{\sqrt{2}}\Big), \nonumber\\\Ket{\Psi_1}=\Big(\frac{\Ket{01}+\Ket{10}}{\sqrt{2}}\Big)
\label{qnm_Eqn23}
\end{eqnarray}
Here, we also use $\alpha_1=\frac{1}{\sqrt{2}}$ and $\alpha_2=\frac{i}{\sqrt{2}}$. The masked quantum state becomes 

\begin{equation}
\Ket{\Psi} =\frac{1}{\sqrt{2}} \Big(\frac{\Ket{00}+\Ket{11}}{\sqrt{2}}\Big) +\frac{i}{\sqrt{2}}\Big(\frac{\Ket{01}+\Ket{10}}{\sqrt{2}}\Big)
\label{qnm_Eqn24}
\end{equation}

\begin{figure*}
\centering
\includegraphics[width=\textwidth]{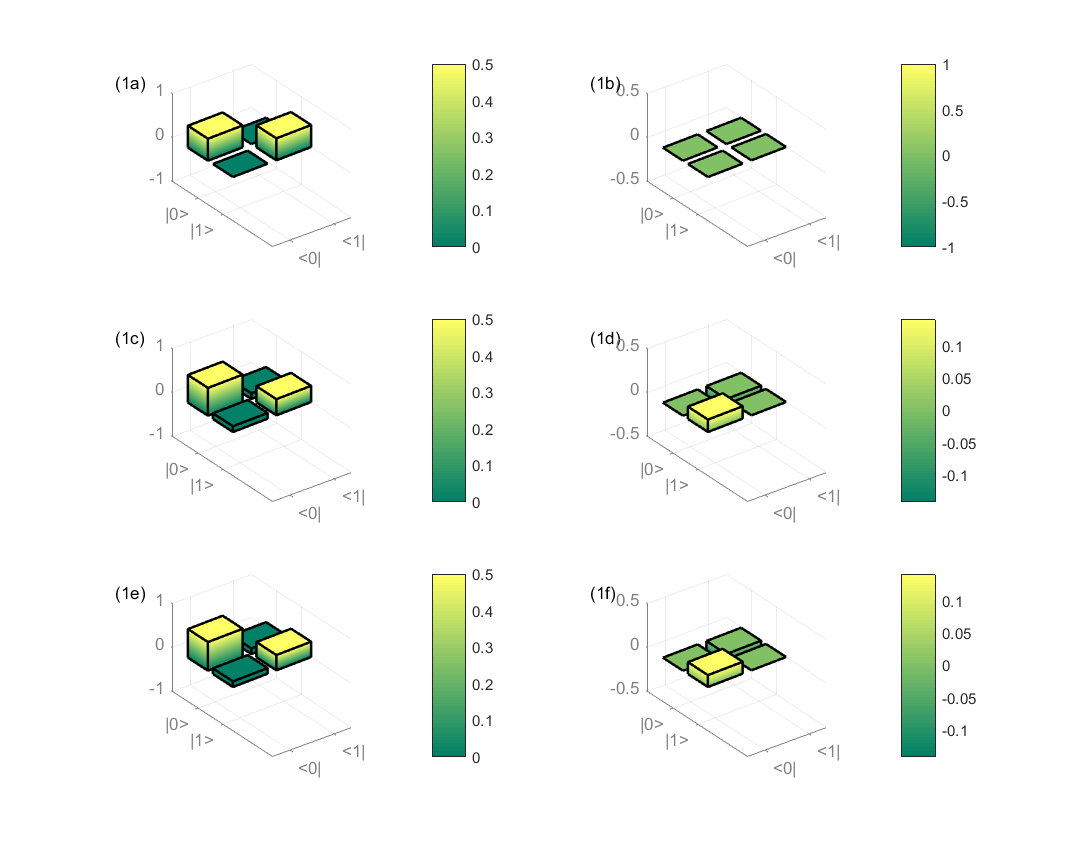}
\caption{1(a) and 1(b) represent the real and imaginary part of the theoretical reduced density matrix for $|\Psi\rangle$ state. 1(c) and 1(d) represent the real and imaginary part of the reconstructed experimental density matrix for $|\Psi_0\rangle$. 1(e) and 1(f) are the real and imaginary part of the experimental density matrix for the state $|\Psi_1\rangle$.}
\label{qnm_Fig4}
\end{figure*}

Theoretical reduced density matrices of these states are

\begin{eqnarray}
Tr_A\ket{\Psi}\big\langle\Psi| = Tr_B\ket{\Psi}\langle\Psi| = \frac{1}{2}(\ket{0}\langle0| + \ket{1}\langle1|)\nonumber\\
Tr_A\ket{\Psi_{0}}\langle\Psi_{0}| = Tr_B\ket{\Psi_{0}}\langle\Psi_{0}| = \frac{1}{2}(\ket{0}\langle0| + \ket{1}\langle1|)\nonumber\\
Tr_A\ket{\Psi_{1}}\langle\Psi_{1}| = Tr_B\ket{\Psi_{1}}\langle\Psi_{1}| = \frac{1}{2}(\ket{0}\langle0| + \ket{1}\langle1|)\nonumber\\
\label{qnm_Eqn25}
\end{eqnarray}

As we can see that the reduced states are the same i.e., $\rho_B = Tr_A\Ket{\Psi_0}\langle\Psi_{0}| = Tr_A\Ket{\Psi_1}\langle\Psi_{1}|$, hence there is no definite information about the initial state. The quantum information is properly masked. Also we know that, it satisfies the masking condition if

\begin{eqnarray}
\alpha_{1}\alpha_{2}^{*}Tr_{X}\Ket{\Psi_{0}}\big\langle\Psi_{1}\big|+\alpha_{1}^{*}\alpha_{2}Tr_{X}\Ket{\Psi_{1}}\big\langle\Psi_{0}\big| = 0 
\label{qnm_Eqn11}
\end{eqnarray}

where, `$X$' represents the sub-systems $A$ or $B$, we take a state whose calculation of partial traces of cross terms gives us,

\begin{eqnarray}
\alpha_{1}\alpha_{2}^{*} Tr_{X}\Ket{\Psi_{0}}\big\langle\Psi_{1}\big| =  \frac{i}{2}\big(\Ket{0}\langle0| + \Ket{1}\langle1|\big)
\label{qnm_Eqn12}
\end{eqnarray}
\begin{eqnarray}
\alpha_{1}^{*}\alpha_{2}Tr_{X}\Ket{\Psi_{1}}\big\langle\Psi_{0}\big| = -\frac{i}{2}\big(\Ket{0}\langle0| + \Ket{1}\langle1|\big)
\label{qnm_Eqn13}
\end{eqnarray}

Using Eq. \eqref{qnm_Eqn12} and Eq. \eqref{qnm_Eqn13} we get,

\begin{eqnarray}
\frac{i}{2}Tr_{X}\Ket{\Psi_{0}}\big\langle\Psi_{1}\big| - \frac{i}{2}Tr_{X}\Ket{\Psi_{1}}\langle\Psi_{0}\big| = 0
\label{qnm_Eqn14}
\end{eqnarray}

which satisfies the masking condition in Eq. \eqref{qnm_Eqn11}. A point is to be noticed that the values of $\alpha_{1}$ and $\alpha_{2}$ can not be arbitrary to fulfill the above condition of masking. There are some restricted values, the above mentioned set of values is one of them. In general case this restriction follows the Eq. \eqref{qnm_Eqn5}. Then we create the above mentioned state on the real chip, ``ibmqx4" with 8192 shots by implementing the quantum gates as shown in Fig. \ref{qnm_Fig2}.

\subsection{Experimental Results}
By coding, we then, construct the density matrices and calculate the partial traces of the density matrices of $\Ket{\Psi_{0}}$, $\Ket{\Psi_{1}}$ and their linear combination state, the input state $\Ket{\Psi}$. The reduced density matrix is calculated firstly by calculating the density matrix of the state by a \textit{Matlab} code. Let, the density matrix is

\begin{eqnarray}
   \rho  = \begin{bmatrix}
   a_{11}&a_{12}&a_{13}&a_{14} \\
   a_{21}&a_{22}&a_{23}&a_{24} \\
   a_{31}&a_{32}&a_{33}&a_{34} \\
   a_{41}&a_{42}&a_{43}&a_{44}
   \end{bmatrix}
   \label{qnm_Eqn15}
\end{eqnarray}
where $a_{ij}$ are the elements of the $\rho$ matrix, i, j = 1, 2, 3, 4 and then the reduced density matrix of a $4\times4$ density matrix of the sub-states $A$ and $B$ will be $2\times2$ matrices and are respectively,

\begin{eqnarray}
\rho_A &=&  \begin{bmatrix}
    a_{11}+a_{22}&a_{13}+a_{24} \\
    a_{31}+a_{42}&a_{33}+a_{44}
   \end{bmatrix}\nonumber\\
\rho_B &=&  \begin{bmatrix}
    a_{11}+a_{33}&a_{12}+a_{34} \\
    a_{21}+a_{43}&a_{22}+a_{44}
   \end{bmatrix}
\label{qnm_Eqn16}
\end{eqnarray}
Now we implement the above masked quantum states $\Psi$, $\Psi_{0}$ and $\Psi_{1}$ and calculate the reduced matrices with the experimental outputs using Eq. \eqref{qnm_Eqn15} and Eq. \eqref{qnm_Eqn16}. The reduced matrices will be $ 2\times2$ matrices, whose diagonal terms are real and off-diagonal terms are imaginary. 

\begin{eqnarray}
\rho_B(\Psi_{0}) &=& \left[\begin{array}{cc}
   0.632&0.130-0.141i \\
   0.130+0.141i&0.368
   \end{array}\right] \nonumber\\
\rho_B(\Psi_{1}) &=& \begin{bmatrix}
   0.652&0.128-0.135i \\
   0.128+0.135i&0.348
   \end{bmatrix}
\label{QNMeq}
\end{eqnarray}

We check the `Distance' between the theoretical and experimental density matrices as well as between the experimental density matrices to show how much they differ from each other. Let us assume $a_T$ is one element of the theoretical density matrix $\rho_T$ and $a_E$ is one element of the experimental density matrix $\rho_E$, then 

\begin{equation}
\hat{D}(\rho_T, \rho_E) = \frac{1}{2}\sum||a_T - a_E||
\label{qnm_Eqn19}
\end{equation}

Then we check the fidelity between theoretical and experimental density matrices to check the density matrices are reconstructed experimentally. Fidelity equals one means the matrices are exactly the same.

\begin{eqnarray}
F(\rho_{T};\rho) = Tr \Big[(\rho_{T})^{\frac{1}{2}}\rho(\rho_{T})^{\frac{1}{2}}\Big]^{\frac{1}{2}}
\label{qnm_Eqn21}
\end{eqnarray}

Here also, we calculate the distances between the density matrices according to the formula in Eq. \eqref{qnm_Eqn19}
\begin{eqnarray}
\hat{D}(\rho_B(\Psi_0);\rho_B(\Psi_1)) = 0.0527 \nonumber\\ 
\hat{D}(\rho_B(\Psi_0);\rho_B(\Psi_{t}) = 0.3239 \nonumber\\
\hat{D}(\rho_B(\Psi_1);\rho_B(\Psi_{t}) = 0.3380 \nonumber\\
\hat{D}(\rho_B(\Psi);\rho_B(\Psi_t) = 0.3006
\label{qnm_Eq}
\end{eqnarray}
where $\Psi_0$ and $\Psi_1$ are represented by Eq. \eqref{qnm_Eqn23}, $\Psi$ is represented by Eq. \eqref{qnm_Eqn24} and $\rho_B(\Psi_0)$, $\rho_B(\Psi_1)$ $\rho_B(\Psi_t)$ are represented by Eq. \eqref{QNMeq}, \eqref{qnm_Eqn25} respectively and the distances with $\Psi_t$ show how perfectly we are doing our work except the first one which depicts how perfectly the masking is happening.

Fidelity can be calculated according to the Eq. \eqref{qnm_Eqn21} to show how much deviation happened in our experiment from theoretical calculations

\begin{eqnarray}
F(\rho_B(\Psi_{t});\rho_B(\Psi_0))=0.9910\nonumber\\
F(\rho_B(\Psi_{t});\rho_B(\Psi_1))=0.9881
\end{eqnarray}
Now we calculate the fidelity between the two practically measured density matrices 
\begin{eqnarray}
F(\rho_B(\Psi_{0});\rho_B(\Psi_1))=0.9997
\end{eqnarray}
This state is also masked with very high fidelity.

\section{Arbitrary Quantum states \label{qnm_Sec4}}
Now, for additional support we create the same states as in Section \ref{qnm_Sec3} but with arbitrary coefficients and see whether it satisfies the masking condition. According to our expectation, as the arbitrary co-efficients don't follow the condition in Eq. \eqref{qnm_Eqn5}
this state should not be masked. We take two states $\Ket\Psi_0 = \alpha\Ket{00} + \beta\Ket{11}$ and $\Ket\Psi_1 = \gamma\Ket{01} + \delta\Ket{10}$ to make 
\begin{eqnarray}
\Ket\Psi = a\big(\Ket{00} + \Ket{11}\big) + b\big(\Ket{01} + \Ket{10}\big)
\label{qnm_Eq26}
\end{eqnarray}
where, $\alpha, \beta$, a, b are chosen arbitrarily.

Instead of using Hadamard gate we use $U$ gate, where angles are taken arbitrarily. 
\begin{eqnarray}
U_3 = \begin{bmatrix}
cos(\frac{\theta}{2})&&-e^{i\lambda}sin(\frac{\theta}{2})\\
e^{i\Phi}sin(\frac{\theta}{2})&&e^{i(\lambda+\Phi)}cos(\frac{\theta}{2})
\end{bmatrix}
\end{eqnarray}

\begin{figure}[H]
\centering
\includegraphics[scale=0.6]{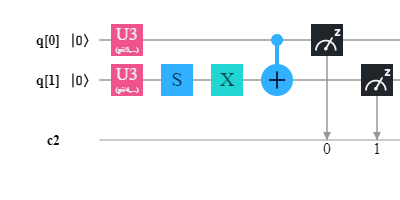}
\caption{Quantum circuit for generating $\Ket{\Psi}=\alpha_1\Ket{\Psi_0}+\alpha_2\Ket{\Psi_1}$, where, $\Ket{\Psi_{0}}=\Big(\frac{\Ket{00}+\Ket{11}}{\sqrt2} \Big)$, and $\Ket{\Psi_{1}}=\Big(\frac{\ket{01}+\Ket{10}}{\sqrt2} \Big)$.}
\label{qnm_Fig5}
\end{figure}

\subsection{Preparation of arbitrary states and calculations}

\begin{figure}[H]
    \centering
    \includegraphics[scale=0.6]{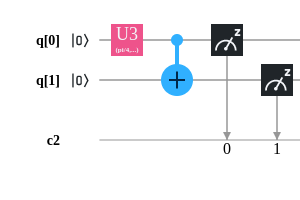}
    \caption{Quantum circuit for generating $\ket{\psi_0 }$}
    \label{qnm_Fig6}
\end{figure}

\begin{figure}[H]
    \centering
    \includegraphics[scale=0.6]{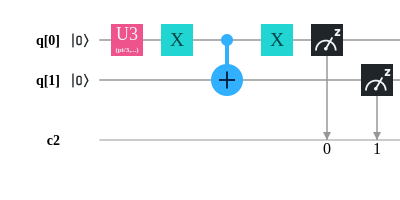}
    \caption{Quantum circuit for generating $\ket{\psi_1 }$}
    \label{qnm_Fig7}
\end{figure}

We prepare arbitrary $\Ket{\Psi_0}$ and $\Ket{\Psi_1}$ states in `ibmq ourense' as shown in Fig. \ref{qnm_Fig6} and Fig. \ref{qnm_Fig7} and we take a unitary gate $U_3$ with arbitrary angles $\pi/4, \pi/4, \pi/5$ respectively to create $\Ket{\Psi_0}$ and arbitrary angles $\pi/3, \pi/4, \pi/5$ respectively to create $\Ket{\Psi_1}$.

First, on the way of making $\Ket{\Psi_0}$ we take the unitary gate, and the gate is represented by matrix as follows

\begin{eqnarray}
U_3(\pi/4, \pi/4, \pi/5) &=& 
\begin{bmatrix}  
   0.9&-0.32-0.24i \\
   0.28+0.28i&0.18+0.9i
\end{bmatrix}
\end{eqnarray}
This gate is applied on first qubit, followed by CNOT gate. Ultimately the state will be 

\begin{eqnarray}
\Ket{\psi_0} = 0.9\Ket{00} + 0.4e^{\frac{i\pi}{4}}\Ket{11}
\end{eqnarray}
Now we trace out the first qubit to get the density matrix

\begin{eqnarray}
\rho_{Bt}({\Psi_0}) &=& 
\begin{bmatrix}  
   0.81&0 \\
   0&0.16
\end{bmatrix}
\end{eqnarray}

Next, on the way of making $\Ket{\Psi_1}$ we take the unitary gate, and the gate is represented by matrix as following

\begin{eqnarray}
U_3(\pi/3, \pi/4, \pi/5) &=& 
\begin{bmatrix}  
   0.9&-0.4-0.3i \\
   0.35+0.35i&0.18+0.9i
\end{bmatrix}
\end{eqnarray}

This gate is applied on first qubit followed by a CNOT gate between two $X$ gates, ultimately the state will be

\begin{eqnarray}
\Ket{\psi_1} = 0.5\Ket{01} + 0.4e^{\frac{i\pi}{4}}\Ket{10}
\end{eqnarray}

We trace out the first qubit to get the density matrix

\begin{eqnarray}
\rho_{Bt}({\Psi_1}) &=& 
\begin{bmatrix}  
   0.25&0 \\
   0&0.81
\end{bmatrix}
\end{eqnarray}
\subsection{Practical Results}

We measure the states $\Ket{\Psi_0}$ and $\Ket{\Psi_1}$ , individually, in IBM quantum computer and put all the values in coding to get the density matrices practically

\begin{eqnarray}
\rho_B(\Psi_{0}) &=& \left[\begin{array}{cc}
   0.836&-0.060-0.017i \\
   -0.060+0.017i&0.164
   \end{array}\right] \nonumber\\
\rho_B(\Psi_{1}) &=& \begin{bmatrix}
   0.284&-0.076+0.027i \\
   -0.076-0.027i&0.716
   \end{bmatrix}
\end{eqnarray}

\subsection{Comparison}
The fidelities between theoretical and practical density matrices to see how perfectly we are doing our work practically
\begin{eqnarray}
F(\rho_{Bt}(\Psi_0);\rho_{B}(\Psi_0)) =  0.9849\nonumber\\
F(\rho_{Bt}(\Psi_1);\rho_{B}(\Psi_1)) =  1
\end{eqnarray}

Now we calculate the distance between practically measured density matrices of two states and also see the fidelity between them to see how likely they are.

According to formula of distance in Eq. \eqref{qnm_Eqn19} the required distance is 

\begin{eqnarray}
\hat{D}(\rho_B(\Psi_0);\rho_B(\Psi_1)) = 0.600 
\label{qnm_Eq}
\end{eqnarray}
And according to formula in Eq. \eqref{qnm_Eqn21} the required  fidelity is
\begin{eqnarray}
F(\rho_{\psi_0};\rho_{\Psi_1}) = 0.8299 
\end{eqnarray}So, compared to previous case, these arbitrary states have too low fidelity and higher distance to feature $\Psi$ to be masked.

\section{Masking into Tripartite states \label{qnm_Sec5}}

So far, in this paper we have dealt with bipartite systems. We have shown the masking of quantum information into orthogonal, and arbitrary quantum states - all are bipartite systems with dimension 2. Li and Wang \cite{qnm_Li2018} have introduced the masking of quantum information into multipartite scenario and claimed that any pure state $S_{d}$ can be masked into a tripartite system except for $d$ = 2, 6 , where `$d$' is the dimension of the system. Later, an interesting fact came out\cite{qnm_Cao2020} that any pure state can be masked into tripartite system and we will see in this section that quantum information can be masked into tripartite system irrespective of the fact that the state follows the restriction on the choice of coefficients i.e. Eq. \eqref{qnm_Eqn5} . Hence, we conclude that \textbf{quantum information can be masked into arbitrary tripartite system unlike the case in bipartite system discussed in section \ref{qnm_Sec4}}.

The only tripartite state that follows the masking condition theoretically maintaining the Eq. \eqref{qnm_Eqn5} is the `Greenberger-Horne-Zeilinger state' \cite{qnm_pan1998}, which is the maximally entangled state among all tripartite states. Trivially the information will be masked into this state . Now, for practical experiment we unfollow the restriction and take arbitrary coefficients by using $U_3$ quantum gate. 

\subsection{Preparation of arbitrary states and calculations}

The state $\Ket{\Phi} = \alpha\ket{000} + \beta\ket{111}$ , where $\alpha$ and $\beta$ are the coefficients which don't follow the restriction. So we apply $U_3$ gate on the first qubit followed by two CNOT gates taking first qubit as control qubit, second qubit as target qubit and second qubit as control qubit, third qubit as target qubit respectively. We make this circuit on the same ‘ibmq ourense’ chip.

\begin{figure}
\centering
\includegraphics[scale=0.5]{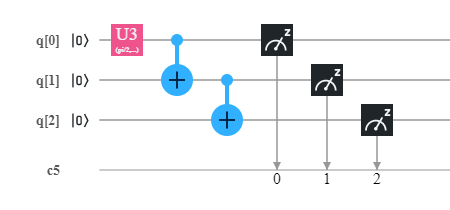}
\caption{Circuit for preparing `Greenberger-Horne-Zeilinger state'\cite{qnm_pan1998}}
\label{qnm_Fig9}
\end{figure}

And we go through the same process. But in this case as the system is tripartite in Hilbert space $H_{A} \otimes H_{B} \otimes H_{C}$,  we get 8$\times$8 density matrix. Tracing the matrix partially we get the partial traces which are 4$\times$4 matrices for subsystem $A$ in Hilbert space $H_{A}$, $B$ in Hilbert space in $H_{B}$ and $C$ in Hilbert space $H_{c}$. The matrices are following:

\begin{widetext}
\begin{eqnarray}
\rho_{A}({\Phi})&=&
0.4693\ket{00}\langle00|+(0.0013-0.068i)\ket{00}\langle01|+(0.0215-0.018i)\ket{00}\langle10|+(-0.0537-0.0021i)\ket{00}\langle11|\nonumber\\&+&(0.0013-0.0068i)\ket{01}\langle00|+0.0392\ket{01}\langle01|+(0.0003-0.0005i)\ket{01}\langle10|+(0.003+0.0045i)\ket{01}\langle11|\nonumber\\&+&(0.0215+0.018i)\ket{10}\langle00|+(0.0003+0.0005i)\ket{10}\langle01|+0.0048\ket{10}\langle10|+(-0.0197+0.0053i)\ket{10}\langle11|\nonumber\\&+&(0.0003+0.0021i)\ket{11}\langle00|+(0.003-0.0045i)\ket{11}\langle01|+(-0.0197-0.0053i)\ket{11}\langle10|+0.4763\ket{11}\langle11|\nonumber\\
\end{eqnarray}
\end{widetext}

\begin{widetext}
\begin{eqnarray}
\rho_{B}({\Phi})&=&
0.4465\ket{00}\langle00|+(0.0048-0.009i)\ket{00}\langle01|+(0.035+0.044i)\ket{00}\langle10|+(-0.0077+0.0107i)\ket{00}\langle11|\nonumber\\&+&(0.0048+0.009i)\ket{01}\langle00|+0.034\ket{01}\langle01|+(0.0022-0.0177i)\ket{01}\langle10|+(-0.0135+0.0105i)\ket{01}\langle11|\nonumber\\&+&(0.0351-0.044i)\ket{10}\langle00|+(0.0022+0.0177i)\ket{10}\langle01|+0.0018\ket{10}\langle10|+(-0.0232+0.0075i)\ket{10}\langle11|\nonumber\\&+&(-0.0077-0.0107i)\ket{11}\langle00|+(-0.0035-0.0105i)\ket{11}\langle01|+(-0.0232-0.0075i)\ket{11}\langle10|+0.4815\ket{11}\langle11|\nonumber\\
\end{eqnarray}
\end{widetext}

\begin{widetext}
\begin{eqnarray}
\rho_{C}({\Phi})&=&
0.4717\ket{00}\langle00|+(0.0152-0.0098i)\ket{00}\langle01|+(0.0223+0.0392i)\ket{00}\langle10|+(-0.0035+0.0398)\ket{00}\langle11|\nonumber\\&+&(0.0152+0.0098i)\ket{01}\langle00|+0.0088\ket{01}\langle01|+(-0.0015-0.0083i)\ket{01}\langle10|+(-0.0007+0.0153i)\ket{01}\langle11|\nonumber\\&+&(0.0223-0.0392i)\ket{10}\langle00|+(-0.0015+0.0083i)\ket{10}\langle01|+0.0368\ket{10}\langle10|+(0.0093-0.0037i)\ket{10}\langle11|\nonumber\\&+&(-0.0035-0.075i)\ket{11}\langle00|+(0.0093-0.0153i)\ket{11}\langle01|+(0.0093+0.0037i)\ket{11}\langle10|+0.4627\ket{11}\langle11|\nonumber\\
\end{eqnarray}
\end{widetext}

\subsection{Practical Results}
Now we measure the fidelity between each partial trace which show how close they are with each other. From \textit{MATLAB} we got the fidelities as follows - 

\begin{eqnarray}
F(\rho_{A}({\Phi});\rho_{B}({\Phi})) = 0.9761\nonumber\\
F(\rho_{B}({\Phi});\rho_{C}({\Phi})) =  0.9564\nonumber\\
F(\rho_{A}({\Phi});\rho_{C}({\Phi})) =  0.9718
\end{eqnarray}

So as expected, the tripartite state which is following the masking condition but not the restriction has high fidelities among the sub-states.

\begin{widetext}
\section{statistical study\label{qnm_Sec6}} 

As one can understand by reading the paper and our introduction as well, this paper mainly aims to verify the `No-Masking theorem' experimentally and on doing that we claim a restricted condition in Eq. \eqref{qnm_Eqn5}. So, our paper focuses on the experimental regime. In this section, we show the statistical errors of our measurements. We take ten trials of our probability measurements, take the average and plot it for different qubits for the systems we considered above. And from the standard deviation for each we plot the error bars for our data.

\begin{table*}[ht]
    \centering
    \scalebox{1.5}{
    \begin{tabular}{|c|c|c|}
    \hline
    &  Orthogonal state & Arbitrary state\\
    \hline
    Mean  & 0.2993 & 0.3092 \\
    \hline
    SD   & 0.0038 & 0.0043 \\
    \hline
    Max  & 0.3050 & 0.3160  \\
    \hline
    Min   & 0.2930 & 0.3070  \\
    \hline
    \end{tabular}}
    \caption{Statistical Parameters in a nutshell for state $\ket{00}$}
    \label{tab:my_label}
\end{table*}
\begin{figure}[H]
\centering
\includegraphics[scale=0.5]{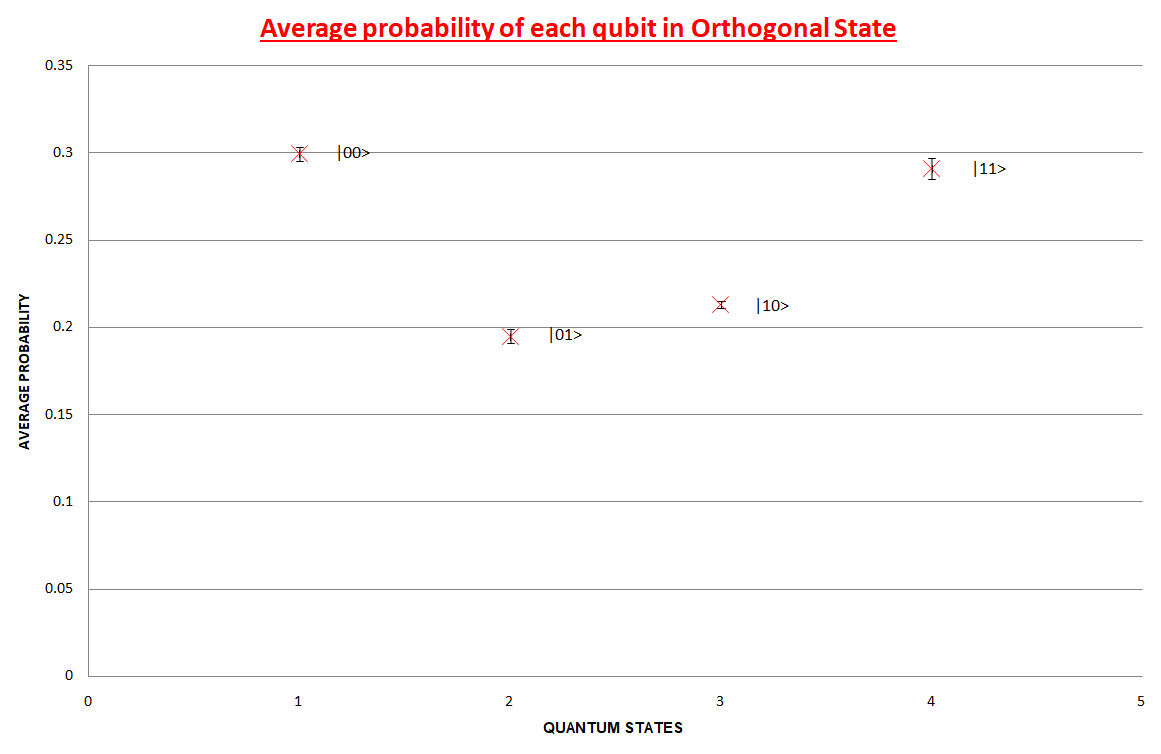}
\caption{Error bars with standard deviation while measuring the states in the circuits to mask information into orthogonal system in Fig. \ref{qnm_Fig2}}
\label{qnm_Fig11}
\end{figure}

\begin{table*}[ht]
    \centering
    \scalebox{1.5}{
    \begin{tabular}{|c|c|c|}
    \hline
    &  Orthogonal state & Arbitrary state\\
    \hline
    Mean  & 0.1948 & 0.1968 \\
    \hline
    SD  & 0.0039 & 0.0064 \\
    \hline
    Max & 0.2010 & 0.2110  \\
    \hline
    Min  & 0.1880 & 0.1900  \\
    \hline
    \end{tabular}}
    \caption{Statistical Parameters for state $\ket{01}$}
    \label{tab:my_label}
\end{table*}

\begin{figure}[H]
\centering
\includegraphics[scale=0.5]{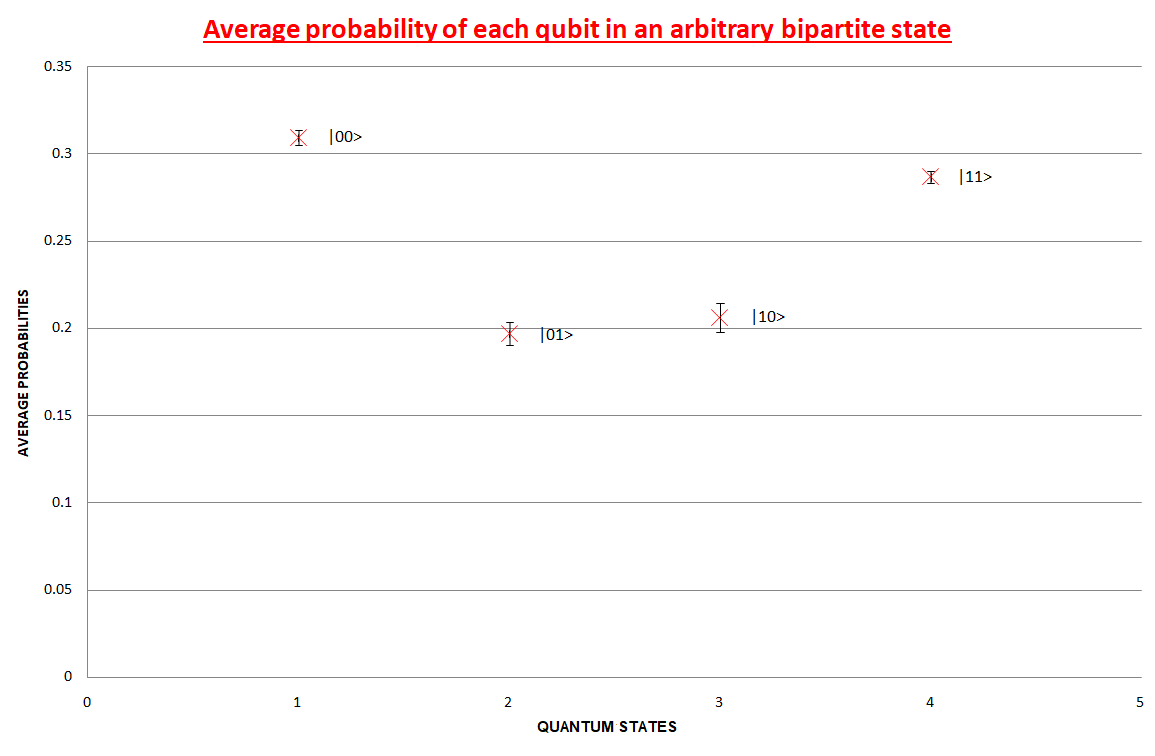}
\caption{Error bars with standard deviation while measuring the states in the circuits to mask information into arbitrary bipartite system in Fig. \ref{qnm_Fig5}}
\label{qnm_Fig12}
\end{figure}
\begin{table*}[ht]
    \centering
    \scalebox{1.5}{
    \begin{tabular}{|c|c|c|c|}
    \hline
    &  Orthogonal state & Arbitrary state\\
    \hline
    Mean & 0.2129 & 0.2060 \\
    \hline
    SD  & 0.0020 & 0.0078 \\
    \hline
    Max  & 0.2160 & 0.2160  \\
    \hline
    Min  & 0.2100 & 0.1940  \\
    \hline
    \end{tabular}}
    \caption{Statistical Parameters for state $\Ket{10}$}
    \label{tab:my_label}
\end{table*}

\begin{figure}[H]
\centering
\includegraphics[scale=0.5]{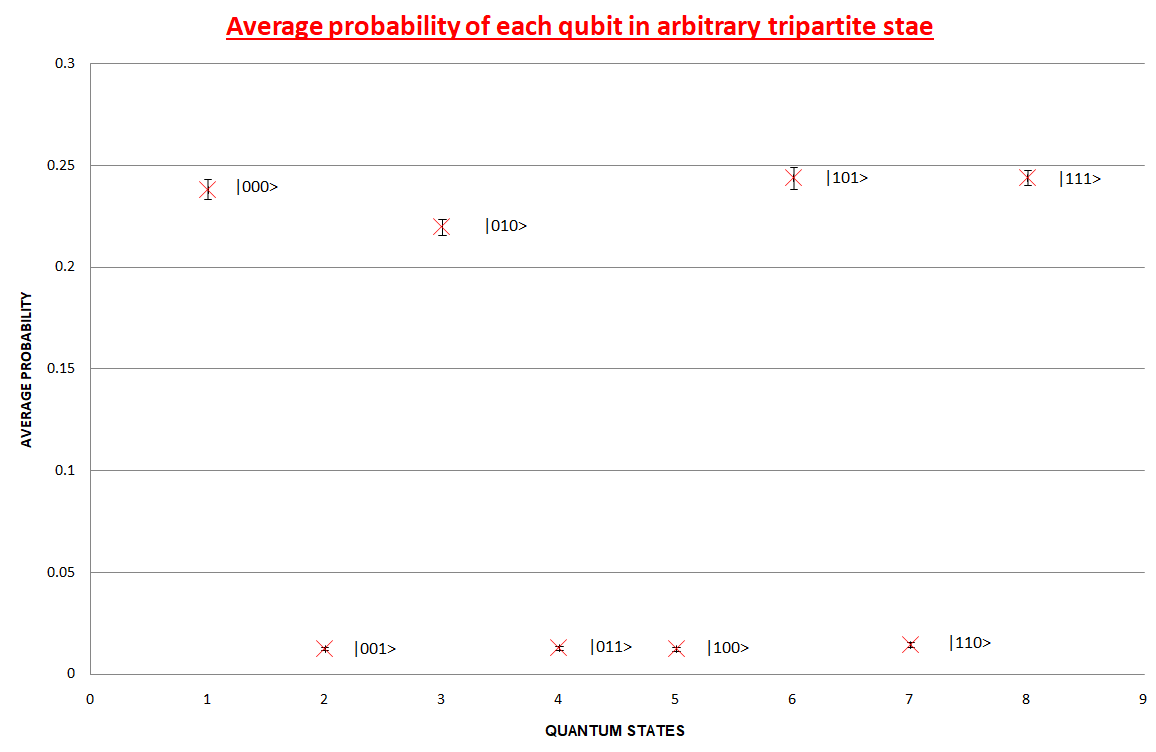}
\caption{Error bars with standard deviation while measuring the states in the circuits to mask information into an arbitrary tripartite system in Fig. \ref{qnm_Fig9}}
\label{qnm_Fig13}
\end{figure}

\begin{table*}[ht]
    \centering
    \scalebox{1.5}{
    \begin{tabular}{|c|c|c|c|}
    \hline
    &  Non orthogonal states & Arbitrary states\\
    \hline
    Mean  & 0.2910 & 0.2866 \\
    \hline
    SD  & 0.0057 & 0.0035 \\
    \hline
    Max & 0.2990 & 0.2920  \\
    \hline
    Min  & 0.2820 & 0.2820  \\
    \hline
    \end{tabular}}
    \caption{Statistical Parameters for state $\Ket{11}$}
    \label{tab:my_label}
\end{table*}

\begin{table*}[ht]
    \centering
    \scalebox{1.5}{
    \begin{tabular}{|c|c|c|c|c|c|c|c|c|}
    \hline
    & $\ket{000}$ & $\ket{001}$ & $\ket{010}$ & $\ket{011}$ & $\ket{100}$ & $\ket{101}$ & $\ket{110}$ & $\ket{000}$ \\
    \hline
    Mean & 0.2381 & 0.0125 & 0.2196 & 0.1291 & 0.0125 & 0.2438 & 0.0147 & 0.2439 \\
    \hline
    SD   & 0.0047 & 0.0009 & 0.0038 & 0.0010 & 0.0009 & 0.0051 & 0.0011 & 0.0036 \\
    \hline
    Max  & 0.2470 & 0.01300 & 0.2230 & 0.0146 & 0.0142 & 0.252 & 0.0172 & 0.249  \\
    \hline
    Min  & 0.2350 & 0.01060 & 0.2140 & 0.0107 & 0.0113 & 0.236 & 0.0136 & 0.236  \\
    \hline
    \end{tabular}}
    \caption{Statistical Parameters for all states in tripartite system}
    \label{tab:my_label}
\end{table*}

\end{widetext}

\section{Discussion \label{qnm_Sec7}}
 Out of all no-go theorems, no-masking theorem is the most current proposed theorem\cite{qnm_ModiPRL2018}. Here, we verify that for some restricted conditions, masking of quantum information is possible. Though it was formerly told that quantum states that contain non-orthogonal states can be masked under certain restricted sets of coefficients, we were surprised to find that under a certain restriction, quantum information containing orthogonal states can be masked too. It means that as long as any quantum information follows that condition or restriction, the information can be masked irrespective of that information being mapped into non-orthogonal or orthogonal state and theoretically Eq. \eqref{qnm_Eqn5} shows this restriction - the main part of our paper. Now to show this practically we performed the experiments for two two-qubit states taking orthogonal basis states into consideration. The states are prepared on the real chip, ``ibmqx4" and found to be masked having high closeness (with fidelity more than 99\%) between the density matrices with more than 98\% fidelitiy (these fidelities show how perfectly the states are prepared in IBM Quantum computer) for orthogonal state in our practical measurements with theoretical calculations respectively. Whereas, for arbitrary quantum states (which do not follow the Eq. \eqref{qnm_Eqn5}) which were prepared on the ``ibmq ourense", we worked out the density matrices for the states practically with more than 99\% fidelities with theoretical calculations (this fidelity shows how perfectly the states are prepared in IBM Quantum computer). However in this case the distance between two density matrices of those same states with arbitrary coefficients is much higher and the fidelity is lower (82\%) as compared to the previous result. According to our expectation these states with arbitrary coefficients cannot be masked. In the last experiment in \ref{qnm_Sec5} we finally show that any tripartite system doesn't follow any restriction and despite using any arbitrary coefficients, the sub- states are similar to each other with high fidelities above 95\% depicting quantum information can be masked into arbitrary tripartite quantum state unlike the case for bipartite system.

So our paper theoretically and experimentally shows that a quantum information can be masked into states (orthogonal) as long as it follows Eq. \eqref{qnm_Eqn5} but if it does not, then those arbitrary quantum information cannot be masked. However, it can be masked into tripartite system without following any restriction. So restriction on the information we are masking is thus shown. For information masking not only the states we are masking into but also the information we are masking must follow the restriction. I hope that our work  not only verifies the newly proposed `No-go theorem' experimentally but also derives another restriction theoretically which will be important for the works in Quantum Masking in the future.

\section*{Acknowledgments}
\label{qlock_acknowledgments}
T.G. and S.S. would like to thank IISER Kolkata for providing hospitality during the course of the project work. B.K.B. acknowledges the support of Institute fellowship provided by IISER Kolkata. The authors acknowledge the support of IBM Quantum Experience. The views expressed are those of the authors and do not reflect the official policy or position of IBM or the IBM Experience team. 


\end{document}